\documentclass[aps, pra, reprint, superscriptaddress,tikz]{revtex4-1}

\usepackage[latin1]{inputenc}
\usepackage{bbding}

\usepackage{tikz}
\usepackage{amsmath,amscd,amsthm,amssymb}
\usepackage{mathrsfs}
\usepackage{graphicx}
\usepackage{epsf,epstopdf}
\usepackage{caption3}
\usepackage[all]{xy}
\usepackage[unicode=true,pdfusetitle, bookmarks=true,bookmarksnumbered=false,bookmarksopen=false, breaklinks=false,pdfborder={0 0 0},backref=false,colorlinks=false] {hyperref}
\hypersetup{ colorlinks,linkcolor=myurlcolor,citecolor=myurlcolor,urlcolor=myurlcolor}
\usepackage{graphics,epstopdf,graphicx, amsthm, amsmath, amssymb, times, braket, color, bm}
\usepackage[up]{subfigure}
\usepackage{cleveref}

\definecolor{myurlcolor}{rgb}{0,0,0.7}

\theoremstyle{plain}
\newtheorem{thm}{\protect\theoremname}

\newtheorem{lem}[thm]{Lemma}
\providecommand{\theoremname}{Theorem}
\newcommand*{\myproofname}{Proof}

\makeatother

\newtheorem*{cor}{Corollary}

\theoremstyle{definition}
\newtheorem{defn}{Definition}

\theoremstyle{remark}
\newtheorem{rem}{Remark}

\begin{document}

 \author{Chunhe Xiong}
\affiliation{Interdisciplinary Center for Quantum Information, School of Physics, Zhejiang University, Hangzhou 310027, China}
\affiliation{School of Computer and Computing Science, Hangzhou City University, Hangzhou 310015, China}

 \author{Sunho Kim}
 \affiliation{School of Mathematical Sciences, Harbin Engineering University, Harbin 150001, China}

 \author{Asutosh Kumar}
\email{asutoshk.phys@gmail.com \Envelope}
\affiliation{P.G. Department of Physics, Gaya College, Magadh University, Rampur, Gaya 823001, India}
 \affiliation{Vaidic and Modern Physics Research Centre, Bhagal Bhim, Bhinmal, Jalore 343029, India}

 \author{Zeyu Chen}
 \affiliation{School of Mathematical Sciences, Zhejiang University, Hangzhou 310027, China}

 \author{Minghui Wu}
\affiliation{School of Computer and Computing Science, Hangzhou City University, Hangzhou 310015, China}

 \author{Junde Wu}
 \email{wjd@zju.edu.cn \Envelope}
 \affiliation{School of Mathematical Sciences, Zhejiang University, Hangzhou 310027, China}

\title{Entanglement as the cross-symmetric part of quantum discord}

\begin{abstract}

In this paper we show that the minimal quantum discord over ``cross-symmetric" state extensions is an entanglement monotone.
In particular, we show that the minimal Bures distance of discord over cross-symmetric extensions is equivalent to the Bures distance of entanglement. At last, we refute a long-held but unstated convention that only contractive distances can be used to construct entanglement monotones by showing that the entanglement quantifier induced by the Hilbert-Schmidt distance, which is not contractive under quantum operations, is also an entanglement monotone.

\end{abstract}
\maketitle

\section{introduction}

Quantum correlation is the fundamental feature that distinguishes quantum world from classical world, and provides remarkable advantages which render quantum information processing much more powerful than classical theory. Entanglement \cite{HorodeckiRMP09} is the most extensive studied form of quantum correlation which plays a fundamental role in quantum tasks \cite{Ekert1991, {Bennett1993}, {Bennett1992}, {Shor1994}}. Quantum correlations beyond entanglement such as quantum discord \cite{Zurek2001A,Vedral2001A,bartlettRMP2007,kavanRMP2012} and quantum coherence \cite{streltsovRMP2017} are also conspicuous and useful in quantum information theory \cite{Hillary2016,Felix2022}. It has been proved that quantum discord, rather than entanglement, is the genuine resource in the deterministic quantum computing with one qubit (DQC1) algorithm \cite{Datta2005,Datta2008A}.

Entanglement and quantum discord have significant differences. In recent years, some remarkable relationships have been investigated between entanglement and discord \cite{Cubbit2003,Koashi2004,Cenlx2011,Adesso2010,Cavalcanti2011,Piani2011,Madhok2011,Streltsov2011,Streltsov2012,Piani2012,Chuan2012}.
Especially, Li and Luo revealed the correspondence between classical states versus separable states \cite{Linan2008}. Moreover, the minimal discord of bipartite quantum state over state extensions is proposed to quantify entanglement \cite{Luo2016,Xiong2022}. Besides, the minimal correlated coherence over symmetric state extensions has been proved to be a good characterization of entanglement \cite{Tan2016,Tan2018}. The characterization and  quantification of entanglement using other quantum correlation or coherence over state extensions is quite different from the existing entanglement measures, viz., entanglement of formation \cite{Bennett1996}, entanglement cost \cite{Bennett1996}, distillable entanglement \cite{Bennett1996c}, relative entropy of entanglement \cite{Vedral1997,Vedral1998}, Bures distance of entanglement \cite{Vedral1997,Vedral1998}, robustness of entanglement \cite{Vidal1999a}, and squashed entanglement \cite{Tucci2002,Christandl2004} which are mostly based on operational meaning, information principles, and mathematics. The interrelationship between these two different approaches of entanglement measures is not clear thus far.

This paper is devoted to the following issues.
At first, we propose a framework to characterize entanglement using the cross-symmetric portion of quantum discord over state extensions. Actually, for a generalized discord measure including entropic discord \cite{Zurek2001A,Vedral2001A}, geometric discord \cite{Dakic2010}, measurement-induced discord \cite{Luo2008B} and correlated coherence \cite{Tan2018}, we prove that the minimal discord over
cross-symmetric state extensions is an entanglement monotone, that is, non-increasing under local operation and classical communications (LOCC). This provides an alternate perspective to understand entanglement from the viewpoint of quantum discord. Moreover, results of Ref.  \cite{Tan2018} are special cases of our work.

Second, we prove that the minimal Bures distance of discord over cross-symmetric state extensions is equivalent to the Bures distance of entanglement.

At last, we show that the non-contractive distances can also be used to quantify entanglement. We show that the minimal Hilbert-Schmidt distance of discord over cross-symmetric state extensions is non-increasing under LOCC operations, even though the Hilbert-Schmidt distance is proved to be not contractive and it is not clear whether this distance can be used to quantify entanglement \cite{Ozawa2000}. Our findings refute a long-held but unstated convention that only contractive distances can be used to construct entanglement monotones.

This paper is structured as follows. In Sec. II we recall
various concepts prerequisite for our study. We introduce the notion of cross-symmetric state extension and characterize entanglement using
quantum discord in Sec. III. We discuss some examples in
Sec. IV and summarize our findings in Sec. V. Appendices
A and B present proofs of Theorem 1 and Lemma 3, respectively.

\section{Preliminaries}


Let $\mathcal{H}=\mathcal{H}_a\otimes\mathcal{H}_b$ and $\mathcal{D}(\mathcal{H})$ be the
set of density matrices on $\mathcal{H}$. A state $\rho_{ab}\in \mathcal{D}(\mathcal{H})$ shared by two parties a and b is called separable if it can be represented as
$
\rho_{ab}=\sum_ip_i\rho_a\otimes\rho_b,
$
where $p_i\ge0, \sum_ip_i=1$ and $\rho_a$, $\rho_b$ are local states for parties $a$ and $b$, respectively. Otherwise, it is called entangled. The set of separable states is denoted by $\mathcal{S}$.

A functional $E$ on $\mathcal{D}(\mathcal{H})$ is called an entanglement monotone if it satisfies (i) faithfulness: $E(\rho)\ge0$, where the equality holds if and only if $\rho\in\mathcal{S}$, and (ii) monotonicity: $E(\rho)\ge E(\Phi(\rho))$ for any LOCC operation $\Phi$.

Moreover, a bipartite state $\rho_{ab}$ is classically correlated if it can be written as
$
 \rho_{ab}=\sum_{i,j}p_{ij}\ket{i}_a\bra{i}\otimes\ket{j}_b\bra{j},
$
where $p_{ij}\ge0$, $\sum_{ij}p_{ij}=1$ and $\{\ket{i}_a\}$ and $\{\ket{j}_b\}$ are orthonormal bases of parties $a$ and $b$, respectively. We denote the set of classical correlated states by $\mathcal{C}\mathcal{C}$. A functional $D$ on $\mathcal{D}(\mathcal{H})$ is called discord measure if it satisfies (D1) faithfulness in $\mathcal{CC}$: $D(\rho)\ge0$, where the equality holds if and only if $\rho\in\mathcal{CC}$, and (D2) invariance under local unitary operations, that is, $D(\rho)=D[(U_a\otimes U_b) \rho (U^{\dagger}_a\otimes U^{\dagger}_b)]$ for any local unitary operations $U_a$ and $U_b$.


\section{entanglement as the minimal discord over cross-symmetric state extensions}

In this section, we show that a number of discord measures can be used to quantify entanglement. We first introduce the notion of cross-symmetric state extension.

\subsection{Cross-symmetric state extension}

\begin{figure}%
\includegraphics[width=3in]{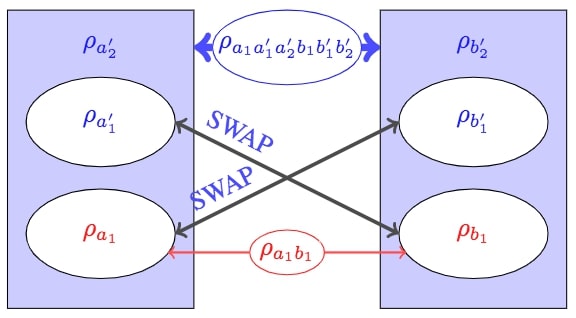}%
\caption{The original state $\rho_{a_1b_1}$ (red color) is shared by party $a$ and $b$. The cross-symmetric extension $\rho_{a_1a^{\prime}_1a^{\prime}_2b_1b^{\prime}_1b^{\prime}_2}$ is invariant under SWAP operation between local subsystem $a_1(b_1)$ and extended local subsystem in $b^{\prime}_1(a^{\prime}_1)$, up to a local unitary operation.}
\label{fig:cross-symmetric-extension}
\end{figure}

\begin{defn}
A cross-symmetric extension (CSE) of a bipartite state $\rho_{a_1b_1}$ is an extension $\rho_{a_1a^{\prime}_1a^{\prime}_2b_1b^{\prime}_1b^{\prime}_2}$ satisfying $\textmd{tr}_{a^{\prime}_1a^{\prime}_2b^{\prime}_1b^{\prime}_2}[\rho_{a_1a^{\prime}_1a^{\prime}_2b_1b^{\prime}_1b^{\prime}_2}]=\rho_{a_1b_1}$
and invariant, up to a local unitary, under the swap operation $\Phi^{a_1\leftrightarrow b^{\prime}_1}_{S}(\Phi^{a^{\prime}_1\leftrightarrow b_1}_{S})$
 between subsystems $a_1(a^{\prime}_1)$ of Alice and $b^{\prime}_1(b_1)$ of Bob, i.e., there exists some unitary operation $U_{a_1a^{\prime}_1a^{\prime}_2}$ such that
\begin{align*}
&\Phi^{a_1\leftrightarrow b^{\prime}_1}_{S}(U_{a_1a^{\prime}_1a^{\prime}_2}\rho_{a_1a^{\prime}_1a^{\prime}_2b_1b^{\prime}_1b^{\prime}_2}U^{\dagger}_{a_1a^{\prime}_1a^{\prime}_2})\nonumber\\
=&U_{a_1a^{\prime}_1a^{\prime}_2}\rho_{a_1a^{\prime}_1a^{\prime}_2b_1b^{\prime}_1b^{\prime}_2}U^{\dagger}_{a_1a^{\prime}_1a^{\prime}_2},
\end{align*}
and similarly for $\Phi^{a^{\prime}_1\leftrightarrow b_1}_{S}$. 
Since the symmetry is ``cross" two subsystems, we call it cross-symmetric extension (see Fig. \ref{fig:cross-symmetric-extension}).
\end{defn}

Furthermore, we claim that the CSE will always exist for a bipartite quantum state. Let us consider a quantum state with spectral decomposition $\rho_{ab}=\sum_i\lambda_i\ket{\psi_i}_{ab}\bra{\psi_i}$. Assume $\ket{\psi_i}_{ab}=\sum_j\sqrt{\mu^i_j}\ket{x^i_j}_a\ket{y^i_j}_b$, then $\ket{\Psi}=\sum_i\sqrt{\lambda_i}\ket{\psi_i}_{ab}\ket{i,i}_{a^{\prime}b^{\prime}}$ is a CSE of $\rho_{ab}$. In fact, one can rewrite $\ket{\Psi}=\sum_{ij}\sqrt{\lambda_i\mu^i_j}\ket{x^i_j,i}_{aa^{\prime}}\ket{y^i_j,i}_{bb^{\prime}}$ and define $U_{aa^{\prime}}=\sum_{i,j}\ket{i,y^i_j}\bra{x^i_j,i}$, then
\begin{align*}
U_{aa^{\prime}}\ket{\Psi}=\sum_{i,j}\sqrt{\lambda_i\mu^i_j}\ket{i,y^i_j}_{aa^{\prime}}\ket{y^i_j,i}_{bb^{\prime}}
\end{align*}
is invariant under swap operation between subsystems $a(a^{\prime})$ and $b^{\prime}(b)$.

\subsection{Quantifying entanglement}

Here we employ the minimal discord over cross-symmetric state extensions to quantify entanglement.

\begin{defn}
For a bipartite state $\rho_{ab}$ and a discord measure $D$, the minimal discord over cross-symmetric extensions is defined as
\begin{align*}
\mathcal{E}_D(\rho_{ab}):=\min_{a^{\prime}b^{\prime}}D(\rho_{aa^{\prime}_1a^{\prime}_2bb^{\prime}_1b^{\prime}_2}),
\end{align*}
where the minimization is performed over all possible CSE of $\rho_{ab}$.
\end{defn}

$\mathcal{E}_D$ being a quantifier of entanglement, it is natural to demand that quantum discord is non-increasing when adding a pure state locally, that is $D(\rho)\ge D(\rho\otimes\ket{0}\bra{0})$. Based on the above definition, we have the following results.

\begin{thm}\label{theorem-elementary-property}
If a discord measure $D$ (faithful in $\mathcal{C}\mathcal{C}$ and local unitary invariant) is non-increasing when adding a pure state ancilla, then $\mathcal{E}_D$ satisfies the following elementary properties:

1. $\mathcal{E}_D(\rho_{ab})\ge0$, the equality holds if and only if $\rho_{ab}$ is separable.

2. $\mathcal{E}_D(\rho_{ab})$ is locally unitary invariant on parties a and b.

3. $\mathcal{E}_D(\rho_{ab})$ is non-increasing under local partial trace in the sense that
\begin{align*}
\mathcal{E}_D(\rho_{ab})\le\mathcal{E}_D(\rho_{aa_1bb_1})
 \end{align*}
 for any state extension $\rho_{aa_1bb_1}$ of $\rho_{ab}$.

4. $\mathcal{E}_D(\rho_{ab})$ is invariant when adding a pure state ancilla,
\begin{align*}
\mathcal{E}_D(\rho_{a_1b_1}\otimes\ket{0}_{b_2}\bra{0})=\mathcal{E}_D(\rho_{a_1b_1}).
\end{align*}

5. $\mathcal{E}_D(\rho_{ab})$ is non-increasing under local operations,
\begin{align*}
\mathcal{E}_D(\rho_{ab})\ge\mathcal{E}_D \left(\sum_iK^i_a\rho_{ab}K^{i\dagger}_a \right).
\end{align*}

6. $\mathcal{E}_D(\rho_{ab})$ is non-increasing under classical communications between parties a and b.

7. $\mathcal{E}_D(\rho_{ab})$ is non-increasing under LOCC operations.
\end{thm}

\begin{proof}
See Appendix \ref{elementary-properties}.
\end{proof}

\begin{rem}
Above results provide a new perspective that the cross-symmetric portion of quantum discord is actually a valid quantifier of entanglement, and this is independent of the particular choice of discord measure. Actually, for a generalized discord measure including entropic discord \cite{Zurek2001A,Vedral2001A}, geometric discord \cite{Dakic2010}, measurement-induced discord \cite{Luo2008B} and correlated coherence \cite{Tan2018}, we prove that the minimal discord over cross-symmetric state extensions is an entanglement monotone.

\end{rem}

\subsection{Geometric discord and measurement-induced discord}

A bivariate function $d$ in the state space $\mathcal{D}(\mathcal{H})$ is called a pseudo-distance if $d(\rho,\sigma)\ge0$ and equality holds iff $\rho=\sigma$. We call $d$ contractive if it satisfies $d(\rho,\sigma)\ge d(\Phi(\rho),\Phi(\sigma))$ for any quantum operation $\Phi$ and $\rho,\sigma\in \mathcal{D}(\mathcal{H})$. In the following we call $d$ simply ``distance".

{\em Geometric discord}--For a bipartite state $\rho_{ab}$ and a contractive distance $d$, the geometric discord is defined as \cite{Vedral1997}
\begin{align*}
D^G_d(\rho_{ab}):=\min_{\sigma_{ab}\in\mathcal{CC}}d(\rho_{ab},\sigma_{ab}),
\end{align*}
where the minimum is taken over all classical correlated states.
It is easy to check that $D^G_d$ is faithful in $\mathcal{CC}$ and local unitary invariant.

{\em Measurement-induced discord}--For a bipartite state $\rho_{ab}$ and a contractive distance $d$, the measurement-induced discord (MID) is defined as \cite{Luo2008B}
\begin{align*}
D^M_d(\rho_{ab}):=\min_{\Pi}d(\rho_{ab},\Pi(\rho_{ab})),
\end{align*}
where the minimum is taken over all local rank-one projective measurements $\{\ket{\alpha_i}_a\bra{\alpha_i}\otimes\ket{\beta_j}_b\bra{\beta_j}\}$, i.e., $\Pi(\rho_{ab})=\sum_{ij}\bra{\alpha_i\beta_j}\rho_{ab}\ket{\alpha_i\beta_j}\ket{\alpha_i}_a\bra{\alpha_i}\otimes\ket{\beta_j}_b\bra{\beta_j}$.

\begin{rem}
The measurement-induced asymmetric discord with relative entropy, Bures distance, Hellinger distance, trace distance, and Hilbert-Schmidt distance have been considered in Refs. \cite{Luo2010,Luo2004,Dakic2010,Nakano2013,Ciccarello2014,Piani2014,Roga2016}. It is easy to check that $D^M_d$ is faithful in $\mathcal{CC}$ and local unitary invariant.
\end{rem}

Many distances like relative entropy, Bures distance, Hellinger distance, and Hilbert-Schmidt distance \cite{note1}, all satisfy
\begin{align}\label{invariant-pure-ancilla}
d(\rho,\sigma)=d(\rho\otimes\ket{0}\bra{0},\sigma\otimes\ket{0}\bra{0}).
\end{align}
Then the induced geometric discords and MIDs are all non-increasing when adding a pure state ancilla \cite{note2}. Based on Theorem \ref{theorem-elementary-property}, we have the following corollary.

\begin{cor}
For each contractive distance $d$ that satisfes (\ref{invariant-pure-ancilla}), the induced quantifiers $\mathcal{E}_{D^G_d}$ and $\mathcal{E}_{D^M_d}$ are entanglement monotones.
\end{cor}

Now let us consider the pure state case. Firstly, assume that $\rho_{aa^{\prime}bb^{\prime}}$ is the optimal extension of $\rho_{ab}$ and $\sigma^{\star}_{aa^{\prime}bb^{\prime}}$ is the closest classical correlated state, then
 \begin{align}\label{lower-bound-by-entanglement}
 \mathcal{E}_{D^G_d}(\rho_{ab})&=D^G_d(\rho_{aa^{\prime}bb^{\prime}})=d(\rho_{aa^{\prime}bb^{\prime}},\sigma^{\star}_{aa^{\prime}bb^{\prime}})\nonumber\\
&\ge\min_{\sigma_{ab}\in\mathcal{CC}}d(\rho_{ab},\sigma_{ab})= E_d(\rho_{ab}),
 \end{align}
where the first inequality holds since the distance $d$ is contractive and the second inequality follows because $\textmd{tr}_{a^{\prime}b^{\prime}}[\sigma^{\star}_{aa^{\prime}bb^{\prime}}]$ is a separable state.

Secondly, for any pure state $\ket{\psi}_{ab}\in \mathcal{D}(\mathcal{H})$, denote the closest classical correlated state by $\sigma^{\star}_{ab}$ such that $D^G_d(\ket{\psi}_{ab})=d(\ket{\psi}_{ab},\sigma^{\star}_{ab})$, then for those distances that satisfy (\ref{invariant-pure-ancilla}), it holds that
\begin{align}\label{upper-bound-by-discord-measure}
\mathcal{E}_{D^G_d}(\ket{\psi}_{ab})&\le D^G_d(\ket{\psi}_{ab}\ket{00}_{a^{\prime}b^{\prime}})\nonumber\\
&\le d(\ket{\psi}_{ab}\ket{00}_{a^{\prime}b^{\prime}},\sigma^{\star}_{ab}\otimes\ket{00}_{a^{\prime}b^{\prime}})\nonumber\\
&=d(\ket{\psi}_{ab},\sigma^{\star}_{ab})=D^G_d(\ket{\psi}_{ab}),
\end{align}
where the inequalities follow from the definition of $\mathcal{E}_{D^G_d}$ and $D^G_d$ and $\ket{\psi}_{ab}\ket{00}_{a^{\prime}b^{\prime}}$ is the cross-symmetric state extension since there exists a local unitary operation $U_{aa^{\prime}}$ such that $U_{aa^{\prime}}(\ket{\psi}_{ab}\ket{00}_{a^{\prime}b^{\prime}})=\ket{\psi}_{a^{\prime}b}\ket{00}_{ab^{\prime}}$.

The above discussion leads to the following result.

\begin{thm}\label{equivalence-with-entanglement-measure}
For those geometric discord measures which reduce to entanglement for pure states, that is, $D^G_d(\ket{\psi}_{ab})=E_d(\ket{\psi}_{ab})$, it holds that\
$$\mathcal{E}_{D^G_d}(\ket{\psi}_{ab})=E_d(\ket{\psi}_{ab}).$$
\end{thm}
\begin{proof}
Combining (\ref{lower-bound-by-entanglement}) and (\ref{upper-bound-by-discord-measure}), for any pure state $\ket{\psi}_{ab}$, we have
\begin{align*}
E_d(\ket{\psi}_{ab})\le\mathcal{E}_{D^G_d}(\ket{\psi}_{ab})\le D^G_d(\ket{\psi}_{ab})=E_d(\ket{\psi}_{ab}).
\end{align*}
\end{proof}

\begin{rem}
For pure states, the geometric discord measures induced by relative entropy, Bures distance and Hellinger distance, all reduce to entanglement monotones. So the induced $\mathcal{E}_{D^G_d}$ is equivalent to the corresponding geometric measure on pure states.
\end{rem}

\section{examples}

In this section, we consider two specific examples to illustrate the above developments.

\subsection{Bures distance case}

The Bures distance is defined as \cite{Bures1969,Uhlmann1976}
\begin{align}\label{Bures-distance}
d_B(\rho,\sigma):=\sqrt{2-2F(\rho,\sigma)},
\end{align}
where $F(\rho,\sigma)$ is the fidelity $F(\rho,\sigma):=tr\sqrt{\sqrt{\sigma}\rho\sqrt{\sigma}}$ between $\rho$ and $\sigma$. Since $F(\rho,\sigma)\in[0,1]$ and is unity iff $\rho=\sigma$, $d_B(\rho,\sigma)$ is nonnegative and vanishes iff $\rho=\sigma$. Moreover, the monotonicity and joint concavity of fidelity \cite{Nielsen10} implies that $d^2_B$ is contractive and jointly convex. Then, the Bures distance of entanglement is defined as \cite{Vedral1997}
$$
E_B(\rho_{ab}):=\min_{\sigma_{ab}\in\mathcal{S}}d^2(\rho_{ab},\sigma_{ab}),
$$
and the convex-roof of Bures distance of entanglement is defined as \cite{wei2003} 
$$E^{cr}_B(\rho_{ab}):=\min_{p_i,\ket{\psi_i}}\sum_ip_iE_B(\ket{\psi_i}),$$
where the minimum is taken over all pure state decompositions $\rho_{ab}=\sum_ip_i\ket{\psi_i}\bra{\psi_i}$.

Obviously, the Bures distance of discord $D^G_B(\rho_{ab}):=\min_{\sigma_{ab}\in\mathcal{CC}}d^2(\rho_{ab},\sigma_{ab})$, similar to the asymmetric case \cite{Aaronson2013,spehner2013A}, is faithful in $\mathcal{C}\mathcal{C}$, local unitary invariant and non-increasing when adding a pure state ancilla. Then the induced entanglement quantifier $\mathcal{E}_{D^G_B}$ is an entanglement monotone, and reduces to the Bures distance of entanglement $E_B(\rho_{ab})$ on pure states.

Next, we show that $\mathcal{E}_{D^G_B}$ is convex and equivalent to $E_B$ on $\mathcal{D}(\mathcal{H})$.

\begin{lem}\label{convexity-induced-Bures-entanglement}
$\mathcal{E}_{D^G_B}$ is convex in the sense that
\begin{align*}
\mathcal{E}_{D^G_B} \left(\sum_ip_i\rho^i_{ab} \right) \le\sum_ip_i\mathcal{E}_{D^G_B}(\rho^i_{ab}),
\end{align*}
where $p_i$ are probabilities and $\rho^i_{ab}$ are bipartite states shared
by parties $a$ and $b$.
\end{lem}

\begin{proof}
See Appendix \ref{convex-of-CSE-Bures-discord}.
\end{proof}

Combining Theorem \ref{equivalence-with-entanglement-measure} and  Lemma \ref{convexity-induced-Bures-entanglement}, we arrive at another theorem below.
\begin{thm}\label{main-theorem}
For $\rho_{ab}\in \mathcal{D}(\mathcal{H})$, the minimal Bures distance of discord over cross-symmetric extensions is equivalent to the Bures distance of entanglement,
\begin{align}
\mathcal{E}_{D^G_B}(\rho_{ab})=E_B(\rho_{ab}).
\end{align}
\end{thm}

\begin{proof}
Since $E_B(\rho_{ab})=E^{cr}_B(\rho_{ab})$ \cite{Streltsov2010}, we just need to show that for each $\rho_{ab}\in \mathcal{D}(\mathcal{H})$, we have
\begin{align*}
E_B(\rho_{ab}) \le \mathcal{E}_{D^G_B}(\rho_{ab}) \le E^{cr}_B(\rho_{ab}),
\end{align*}
where the first inequality follow from Eq.~(\ref{lower-bound-by-entanglement}) and we derive the  second inequality below.

Actually, for any mixed state with pure state decomposition $\rho_{ab}=\sum_ip_i\ket{\psi_i}_{ab}\bra{\psi_i}$, Theorem \ref{convexity-induced-Bures-entanglement} tells us that
\begin{align*}
\mathcal{E}_{D^G_B}(\rho_{ab})\le\sum_ip_i\mathcal{E}_{D^G_B}(\ket{\psi_i})=\sum_ip_iE_B(\ket{\psi_i}),
\end{align*}
where the equality follows from Theorem \ref{equivalence-with-entanglement-measure}.
Taking the minimum over all pure state decompositions,
\begin{align*}
\mathcal{E}_{D^G_B}(\rho_{ab})\le E^{cr}_B(\rho_{ab}).
\end{align*}
\end{proof}

\subsection{Hilbert Schmidt distance case}

It is well known that the Hilbert-Schmidt distance is not contractive and it is not clear whether this distance can be used to quantify entanglement \cite{Ozawa2000}. Here we show that the minimal Hilbert-Schmidt distance of discord over cross-symmetric state extensions is an entanglement monotone.

Recall that the Hilbert-Schmidt distance is
$
d_{HS}(\rho,\sigma)=\sqrt{\textmd{tr}(\rho^2+\sigma^2-2\rho\sigma)}
$
and the corresponding discord measure is defined by $D^G_{HS}(\rho):=d_{HS}(\rho,\sigma)$.
By definition, $D^G_{HS}$ is faithful in $\mathcal{C}\mathcal{C}$ and local unitary invariant. Obviously, $d_{HS}(\rho,\sigma)=d_{HS}(\rho\otimes\ket{0}\bra{0},\sigma\otimes\ket{0}\bra{0})$, then $D^G_{HS}$ is contractive when adding a pure state ancilla \cite{note2}. Let us denote the  entanglement measure corresponding to $D^G_{HS}$ by $\mathcal{E}_{D^G_{HS}}$.

Based on Theorem \ref{theorem-elementary-property}, we have the following result.
\begin{cor}
For the Hilbert-Schmidt distance $d_{HS}$, the induced entanglement quantifier $\mathcal{E}_{D^G_{HS}}$ is an entanglement monotone.
\end{cor}

\begin{rem}
Although Hilbert Schmidt distance is not contractive, the induced quantifier $\mathcal{E}_{D^G_{HS}}$ is contractive under LOCC operations.
\end{rem}

\section{Conclusion}

In summary, we constructed a large class of entanglement measures using the quantum discord measures. Actually, we showed that the minimal discord over cross-symmetric state extensions is non-increasing under LOCC provided the discord measure is invariant when adding pure state ancilla. This result establishes an interesting connection between entanglement and discord.

A common way to construct entanglement monotone is to choose a contractive distance such as relative entropy, Bure distance, etc. Then the entanglement monotone is defined as the minimum distance of a quantum state from a set of separable states. This approach obviously limits the means of quantifying entanglement. For instance, the Hilbert-Schmidt distance is not suitable to quantify entanglement. In this work, we proposed a framework to quantify entanglement using the minimal discord over cross-symmetric state extensions. The conditions ``over state extension" and non-increasing when adding a pure state ancilla ensure that the quantifier is contractive under local operations, and the invariance under SWAP operations ensures that the quantifier is contractive under classical communications. The framework presented here greatly relaxes the constraint of contractive distance, which sheds new light on the quantitative study of entanglement.

\begin{acknowledgments}
This project is supported by the National Natural Science Foundation of China (Grants No. 12201555 and No. 12050410232, 12031004, 12271474) and  the Postdoctoral Science Foundation of China (Grant No. 2021M702864).
\end{acknowledgments}

\appendix

\section{Proof of Theorem \ref{theorem-elementary-property}}\label{elementary-properties}

The proofs are presented in the same order as they are mentioned in the main text.

{\em Property} 1 (Faithfulness.) $\mathcal{E}_D(\rho_{ab})\ge0$, the equality holds if and only if $\rho_{ab}$ is separable.
\begin{proof}
Discord is nonnegative over valid quantum states. Therefore,
$\mathcal{E}_D(\rho_{ab})$ being discord of the extended state, must also be nonnegative.

A separable bipartite state $\rho_{ab}$ can be written as $\rho_{ab}=\sum_ip_i\rho^i_a\otimes\sigma^i_b$.  Assume $\rho^i_a=\sum_jq_{i,j}\ket{a_{i,j}}\bra{a_{i,j}}$ and $\sigma^i_b=\sum_kr_{i,k}\ket{b_{i,k}}\bra{b_{i,k}}$. Then, up to a relabelling of variables, the separable state can be written as a convex sum of separable pure states of the form $\rho_{ab}=\sum_ls_l\ket{a_l}\bra{a_l}\otimes\ket{b_l}\bra{b_l}$. Therefore, $\rho_{aa^{\prime}a^{\prime}_1bb^{\prime}b^{\prime}_1}=\sum_ls_l\ket{a_l,b_l,l}_{aa^{\prime}a^{\prime}_1}\bra{a_l,b_l,l}\otimes\ket{b_l,a_l,l}_{bb^{\prime}b^{\prime}_1}\bra{b_l,a_l,l}$ is a CSE of $\rho_{ab}$ and $\mathcal{E}_D(\rho_{ab})\le D(\rho_{aa^{\prime}a^{\prime}_1bb^{\prime}b^{\prime}_1})=0$.

Conversely, $\mathcal{E}_D(\rho_{ab})=0$ implies that there exists an extension for which $D(\rho_{aa^{\prime}bb^{\prime}})=0$, i.e., $\rho_{aa^{\prime}bb^{\prime}}=\sum_it_i\ket{\alpha_i}_{aa^{\prime}}\bra{\alpha_i}\otimes\ket{\beta_i}_{bb^{\prime}}\bra{\beta_i}$. Tracing out the subsystem $a^{\prime}b^{\prime}$ will lead to the decomposition of the form $\rho_{ab}=\sum_it_i\ket{a_i}_a\bra{a_i}\otimes\ket{b_i}_b\bra{b_i}$ which is separable.

In conclusion, $\mathcal{E}_D(\rho_{ab})$ must be strictly positive for entangled states since it is nonnegative and vanishes iff $\rho_{ab}$ is separable.
\end{proof}

{\em Property} 2 (Invariance under local unitaries.) $\mathcal{E}_D(\cdot)$ is locally unitary invariant on parties $a$ and $b$ in the sense that
\begin{align*}
\mathcal{E}_D \left[(U_a\otimes U_b)\rho_{ab}(U^{\dagger}_a\otimes U^{\dagger}_b) \right]=\mathcal{E}_D(\rho_{ab}).
\end{align*}
\begin{proof}
This property follows from the fact that $D(\rho_{ab})$ is invariant under local unitary operations.
\end{proof}

{\em Property} 3 (Contraction under local partial trace.) $\mathcal{E}_D(\cdot)$ is nonincreasing under local partial trace in the sense that
\begin{align*}
\mathcal{E}_D(\rho_{ab})\le \mathcal{E}_D(\rho_{aa_1bb_1})
\end{align*}
 for any state extension $\rho_{aa_1bb_1}$ of $\rho_{ab}$.

\begin{proof}
Since any CSE $\rho_{aa_1a^{\prime}a^{\prime}_1a^{\prime}_2bb_1b^{\prime}b^{\prime}_1b^{\prime}_2}$ of $\rho_{aa_1bb_1}$ is necessarily a state extension of the reduced state $\rho_{ab}=tr_{a_1b_1}\rho_{aa_1bb_1}$ which satisfies
\begin{align*}
&\Phi^{aa_1\leftrightarrow b^{\prime}b^{\prime}_1}_S(U_{aa_1a^{\prime}a^{\prime}_1a^{\prime}_2}\rho_{aa_1a^{\prime}a^{\prime}_1a^{\prime}_2bb_1b^{\prime}b^{\prime}_1b^{\prime}_2}U^{\dagger}_{aa_1a^{\prime}a^{\prime}_1a^{\prime}_2})\nonumber\\
=&U_{aa_1a^{\prime}a^{\prime}_1a^{\prime}_2}\rho_{aa_1a^{\prime}a^{\prime}_1a^{\prime}_2bb_1b^{\prime}b^{\prime}_1b^{\prime}_2}U^{\dagger}_{aa_1a^{\prime}a^{\prime}_1a^{\prime}_2},
\end{align*}
for some local unitary $U_{aa_1a^{\prime}a^{\prime}_1a^{\prime}_2}$ and a similar equality for $\Phi^{a^{\prime}a^{\prime}_1\leftrightarrow bb_1}_S$, then we have
\begin{align*}
\mathcal{E}_D(\rho_{aa_1bb_1})=&\min_{\{\rho_{aa_1bb_1},\Psi^{a^{\prime}a^{\prime}_1\leftrightarrow bb_1}_S\}} D(\rho_{aa_1a^{\prime}a^{\prime}_1a^{\prime}_2bb_1b^{\prime}b^{\prime}_1b^{\prime}_2})\nonumber\\
\ge&\min_{\{\rho_{ab},\Psi^{a^{\prime}\leftrightarrow b}_S\}} D(\rho_{aa_1a^{\prime}a^{\prime}_1a^{\prime}_2bb_1b^{\prime}b^{\prime}_1b^{\prime}_2})\ge\tilde{\mathcal{E}}_D(\rho_{ab}),
\end{align*}
where the first minimum is taken over all  $\rho_{aa_1a^{\prime}a^{\prime}_1a^{\prime}_2bb_1b^{\prime}b^{\prime}_1b^{\prime}_2}$ which is the extension of $\rho_{aa_1bb_1}$ and invariant under swap operation between subsystem $aa_1(a^{\prime}a^{\prime}_1)$ and $b^{\prime}b^{\prime}_1(bb_1)$, while the second minimum is taken over all $\rho_{aa_1a^{\prime}a^{\prime}_1a^{\prime}_2bb_1b^{\prime}b^{\prime}_1b^{\prime}_2}$ that is the extension of $\rho_{ab}$ and invariant under swap operation between subsystem $a(a^{\prime})$ and $b^{\prime}(b)$. Therefore, the chain of inequalities holds by definition.
\end{proof}

{\em Property} 4. $\mathcal{E}_D(\cdot)$ is invariant when adding a pure state ancilla,
\begin{align}\label{invariant-adding-pure-ancilla}
\mathcal{E}_D(\rho_{a_1b_1}\otimes\ket{0}_{b_2}\bra{0})=\mathcal{E}_D(\rho_{a_1b_1}).
\end{align}

\begin{proof}
Note that the CSE of $\rho_{a_1b_1}\otimes\ket{0}_{b_2}\bra{0}$ has the form $\rho_{a_1a^{\prime}_1a^{\prime}_2a^{\prime}_3b_1b^{\prime}_1b^{\prime}_3}\otimes\ket{0}_{b_2}\bra{0}$. 
Therefore,
\begin{align*}
\mathcal{E}_D(\rho_{a_1b_1})&=\min_{\{\rho_{a_1b_1},\Psi^{a_1\leftrightarrow b^{\prime}_1}_S\}} D(\rho_{a_1a^{\prime}_1a^{\prime}_3b_1b^{\prime}_1b^{\prime}_3})\nonumber\\
&\ge\min_{\{\rho_{a_1b_1},\Psi^{a_1\leftrightarrow b^{\prime}_1}_S\}} D(\rho_{a_1a^{\prime}_1a^{\prime}_3b_1b^{\prime}_1b^{\prime}_3}\otimes\ket{00}_{a^{\prime}_2b_2}\bra{00})\nonumber\\
&\ge \mathcal{E}_D(\rho_{a_1b_1}\otimes\ket{0}_{b_2}\bra{0}),
\end{align*}
where the minimum is taken over all $\rho_{aa_1a^{\prime}a^{\prime}_1a^{\prime}_2bb_1b^{\prime}b^{\prime}_1b^{\prime}_2}$ which is the extension of $\rho_{ab}$ and invariant under SWAP operation between subsystem $a(a^{\prime})$ and $b^{\prime}(b)$. Using property 3 and the fact that $\rho_{a_1b_1}\otimes\ket{0}_{b_2}\bra{0}$ is a state extension of $\rho_{ab}$, $\mathcal{E}_D(\rho_{a_1b_1})=\mathcal{E}_D(\rho_{a_1b_1}\otimes\ket{0}_{b_2}\bra{0})$.
\end{proof}


{\em Property} 5 (Contraction under local operations.) $\mathcal{E}_D(\cdot)$ is nonincreasing under local operations,
\begin{align*}
\mathcal{E}_D(\rho_{ab})\ge \mathcal{E}_D \left(\sum_iK^i_a\rho_{ab}K^{i\dagger}_a \right).
\end{align*}

\begin{proof}
Due to Stinespring representation \cite{stinespring1955}, the local operation can be realized by adding a pure state ancilla, followed by a global unitary operation and tracing out the ancilla system, i.e., $\sum_iK^i_a\rho_{ab}K^{i\dagger}_a=tr_{a_1}U_{aa_1}(\rho_{ab}\otimes\ket{0}_{a_1}\bra{0})U^{\dagger}_{aa_1}$. Therefore, one has
\begin{align*}
\mathcal{E}_D (\rho_{ab})&=\mathcal{E}_D \left(\rho_{ab}\otimes\ket{0}_{a_1}\bra{0} \right)\\
&=\mathcal{E}_D \left(U_{aa_1}\rho_{ab}\otimes\ket{0}_{a_1}\bra{0}U^{\dagger}_{aa_1} \right)\\
&\ge\mathcal{E}_D \left(\sum_iK^i_a\rho_{ab}K^{i\dagger}_a \right),
\end{align*}
where the first equality follows from property 4 and the last inequality follows from property 3 because the state in the second line is an extension of $\sum_iK^i_a\rho_{ab}K^{i\dagger}_a$.

\end{proof}

\begin{widetext}
{\em Property} 6 (Contraction under classical communications.)  Consider a bipartite state $\rho_{ab}$ shared by Alice and Bob, and a classical register $b_1$ which stores classical information in Bob's side. The resultant state, after Bob performs a local measurement, is $\sum_ip_i\rho^i_{ab}\otimes\ket{i}_{b_1}\bra{i}$.
Then $\mathcal{E}_D$ is non-increasing in the sense that
\begin{align*}
\mathcal{E}_D \left(\sum_ip_i\rho^i_{ab}\otimes\ket{i}_{b_1}\bra{i} \right)\ge \mathcal{E}_D \left(\sum_ip_i\rho^i_{ab}\otimes\ket{i}_{M_a}\bra{i}\otimes\ket{i}_{M_b}\bra{i} \right),
\end{align*}
where $\sum_ip_i\rho^i_{ab}\otimes\ket{i}_{M_a}\bra{i}\otimes\ket{i}_{M_b}\bra{i}$ is equivalent to the state after Bob sends classical information stored in the computational basis of the register $b_1$ to Alice.

\begin{proof}
The outline of the proof is similar to that in Ref. \cite{Tan2018}.
Assume that $\sigma^{\star}_{aa^{\prime}a^{\prime}_1a^{\prime}_2bb_1b^{\prime}b^{\prime}_2}$ is the optimal CSE of $\sum_ip_i\rho^i_{ab}\otimes\ket{i}_{b_1}\bra{i}$. By definition, there  exists a local unitary that Alice can perform such that
\begin{align*}
U_{aa^{\prime}a^{\prime}_1a^{\prime}_2}\sigma^{\star}_{aa^{\prime}a^{\prime}_1a^{\prime}_2bb_1b^{\prime}b^{\prime}_2}U^{\dagger}_{aa^{\prime}a^{\prime}_1a^{\prime}_2}=\Phi^{a^{\prime}_1\leftrightarrow b_1}_S(U_{aa^{\prime}a^{\prime}_1a^{\prime}_2}\sigma^{\star}_{aa^{\prime}a^{\prime}_1a^{\prime}_2bb_1b^{\prime}b^{\prime}_2}U^{\dagger}_{aa^{\prime}a^{\prime}_1a^{\prime}_2}).
\end{align*}

Suppose we add registers $M_a,M^{\prime}_{a}$ and $M_b,M^{\prime}_{b}$, initialized in the state $\ket{0,0}_{M_aM^{\prime}_{a}}$ and $\ket{0,0}_{M_bM^{\prime}_{b}}$, and locally copy the
classical information on registers $a^{\prime}_1$ and $b_1$ via CNOT operations $\mathcal{U}^{a^{\prime}_1M_a}_C$ and $\mathcal{U}^{b_1M_b}_C$ with $\mathcal{U}^{ab}_C(\cdot)=U^{ab}_C(\cdot)U^{ab\dagger}_C$ and $U^{ab}_C=\sum_{ij}\ket{i,i+j(\mathrm{mod}~d_{\mathcal{H}_b})}\bra{i,j}$. It can be verified that $\mathcal{U}^{ab}_C$ is unitary and satisfies $U^{ab}_C\ket{i,0}_{ab}=\ket{i,i}_{ab}$.
Assume that this unitary operation is already performed and included in the definition of $\sigma^{\star}_{aa^{\prime}a^{\prime}_1a^{\prime}_2bb_1b^{\prime}b^{\prime}_2}$. Then the invariance under SWAP operations implies that
\begin{align}
&\mathcal{U}^{a^{\prime}_1M_a}_C\circ\mathcal{U}^{b_1M_b}_C \left(\ket{0,0}_{M_aM^{\prime}_{a}}\bra{0,0}\otimes\sigma^{\star}_{aa^{\prime}a^{\prime}_1a^{\prime}_2bb_1b^{\prime}b^{\prime}_2}\otimes\ket{0,0}_{M_bM^{\prime}_{b}}\bra{0,0} \right)\\
=&\Phi^{a^{\prime}_1\leftrightarrow b_1}_S\circ\Phi^{a^{\prime}_1\leftrightarrow b_1}_S\circ\mathcal{U}^{a^{\prime}_1M_a}_C\circ\mathcal{U}^{b_1M_b}_C(\ket{0,0}_{M_aM^{\prime}_{a}}\bra{0,0}\otimes\Phi^{a^{\prime}_1\leftrightarrow b_1}_S(\sigma^{\star}_{aa^{\prime}a^{\prime}_1a^{\prime}_2bb_1b^{\prime}b^{\prime}_2})\otimes\ket{0,0}_{M_bM^{\prime}_{b}}\bra{0,0})\\
=&\Phi^{a^{\prime}_1\leftrightarrow b_1}_S\circ\mathcal{U}^{b_1M_a}_C\circ\mathcal{U}^{a^{\prime}_1M_b}_C(\ket{0,0}_{M_aM^{\prime}_{a}}\bra{0,0}\otimes\sigma^{\star}_{aa^{\prime}a^{\prime}_1a^{\prime}_2bb_1b^{\prime}b^{\prime}_2}\otimes\ket{0,0}_{M_bM^{\prime}_{b}}\bra{0,0}),
\end{align}
where $\mathcal{U}^{ab}_C(\rho_{ab})=U^{ab}_C\rho_{ab}U^{ab\dagger}_C$. The last equality follows from the fact that $\Phi^{a\leftrightarrow b}_S(\rho_{ab})=U^{a\leftrightarrow b}_S\rho_{ab}U^{a\leftrightarrow b\dagger}_S$, $U^{a\leftrightarrow b}_S=U^{a\leftrightarrow b\dagger}_S$ and $U^{a\leftrightarrow b}_SU^{bc}_CU^{a\leftrightarrow b\dagger}_S=U^{ac}_C$.

It can be verified that Eq.~(A2) is a CSE of $\sum_i\ket{i}_{M_a}\bra{i}\otimes \rho^i_{ab}\otimes\ket{i}_{M_b}\bra{i}$,  which is equivalent
to the state after Bob sends classical information to Alice. To see this, let us perform the following partial trace on Eq.~(A4):
\begin{align}
&\textmd{tr}_{a^{\prime}a^{\prime}_1a^{\prime}_2b_1b^{\prime}b^{\prime}_2} \left(\Phi^{a^{\prime}_1\leftrightarrow b_1}_S\circ\mathcal{U}^{b_1M_a}_C\circ\mathcal{U}^{a^{\prime}_1M_b}_C(\ket{0,0}_{M_aM^{\prime}_{a}}\bra{0,0}\otimes\sigma^{\star}_{aa^{\prime}a^{\prime}_1a^{\prime}_2bb_1b^{\prime}b^{\prime}_2}\otimes\ket{0,0}_{M_bM^{\prime}_{b}}\bra{0,0} \right)\\
=&\textmd{tr}_{a^{\prime}a^{\prime}_1a^{\prime}_2b_1b^{\prime}b^{\prime}_2}(\mathcal{U}^{b_1M_a}_C\circ\mathcal{U}^{a^{\prime}_1M_b}_C(\ket{0,0}_{M_aM^{\prime}_{a}}\bra{0,0}\otimes\sigma^{\star}_{aa^{\prime}a^{\prime}_1a^{\prime}_2bb_1b^{\prime}b^{\prime}_2}\otimes\ket{0,0}_{M_bM^{\prime}_{b}}\bra{0,0}),
\end{align}
where the equality follows from the fact that SWAP does not affect the partial trace, i.e., $\textmd{tr}_{ab}\circ\Phi^{a\leftrightarrow b}_S=\textmd{tr}_{ab}$. Note that in Eq.~(A6), register $M_a$ contains a copy of the information in register $b_1$ in the computational basis, even though Alice just performed a local operation in Eq.~(A2). We also see in Eq.~(A2) that register $M_b$ just contains a copy of classical information in register $b_1$. Therefore, this implies that Eq.~(A2) is in fact a CSE of the state $\sum_ip_i\ket{i}_{M_a}\bra{i}\otimes \rho^i_{ab}\otimes\ket{i}_{M_b}\bra{i}$.

In conclusion,
\begin{align*}
&\mathcal{E}_D \left(\sum_ip_i\rho^i_{ab}\otimes\ket{i}_{b_1}\bra{i} \right)=D \left(\sigma^{\star}_{aa^{\prime}a^{\prime}_1a^{\prime}_2bb_1b^{\prime}b^{\prime}_2} \right) = D \left(\ket{0,0}_{M_aM^{\prime}_{a}}\bra{0,0}\otimes\sigma^{\star}_{aa^{\prime}a^{\prime}_1a^{\prime}_2bb_1b^{\prime}b^{\prime}_2}\otimes\ket{0,0}_{M_bM^{\prime}_{b}}\bra{0,0} \right)\\
=&D \left[\mathcal{U}^{a^{\prime}_1M_a}_C\circ\mathcal{U}^{b_1M_b}_C \left(\ket{0,0}_{M_aM^{\prime}_{a}}\bra{0,0}\otimes\sigma^{\star}_{aa^{\prime}a^{\prime}_1a^{\prime}_2bb_1b^{\prime}b^{\prime}_2}\otimes\ket{0,0}_{M_bM^{\prime}_{b}}\bra{0,0} \right) \right] \ge \mathcal{E}_D \left(\sum_ip_i\ket{i}_{M_a}\bra{i}\otimes \rho^i_{ab}\otimes\ket{i}_{M_b}\bra{i} \right),
\end{align*}
where the first three equalities
follow from the definition and properties 2, 3, 4, and the last inequality follows from the fact that Eq.~(A2) is a CSE of the last state. This is already sufficient for us to prove that $\mathcal{E}_D$ cannot increase under classical communication.

\end{proof}

\end{widetext}

{\em Property} 7 (Contraction under LOCC.) $\mathcal{E}_D(\cdot)$ is nonincreasing under LOCC operations.

\begin{proof}
The local measurement can be implemented by adding a local pure state ancilla followed by evolving through a unitary operation and performing the von Neumann measurements. And any LOCC operation consists of a series of local quantum operations and classical communication. We prove that $\mathcal{E}_D$ is contractive at each step.

Firstly, suppose Bob (represented by the subsystem $b$) communicates classical information
to Alice (represented by subsystem $a$). By Naimark's theorem, Bob's local quantum operation can be realized by adding ancillary subsystems $b_1b_2$ in the initial pure state $\ket{0}_{b_1}\bra{0}\otimes\ket{0}_{b_2}\bra{0}$, followed by a unitary operation on $bb_1b_2$. Let subsystem $b_1$ contains all the classical information after the unitary is performed and $b_2$ is traced out.

Based on the above discussion, one has
\begin{align*}
\mathcal{E}_D(\rho_{ab})&\ge \mathcal{E}_D \left(\rho_{ab}\otimes\ket{0}_{b_1}\bra{0}\otimes\ket{0}_{b_2}\bra{0} \right)\\
=&\mathcal{E}_D \left(U_{bb_1b_2}\rho_{ab}\otimes\ket{0}_{b_1}\bra{0}\otimes\ket{0}_{b_2}\bra{0}U^{\dagger}_{bb_1b_2} \right)\\
\ge&\mathcal{E}_D \left(\sum_iK^i_b\rho_{ab}K^{i\dagger}_b\otimes\ket{i}_{b_1}\bra{i} \right),
\end{align*}
where the chain of inequalities follow from properties 2,3,4, and the observation that the state in the penultimate parenthesis is an extension of the last one.

Secondly, property 6 guarantees that classical communication does not increase $\mathcal{E}_D$, that is,
\begin{align*}
&\mathcal{E}_D \left(\sum_iK^i_b\rho_{ab}K^{i\dagger}_b\otimes\ket{i}_{b_1}\bra{i} \right)\\
\ge&\mathcal{E}_D \left(\sum_i\ket{i}_{a_1}\bra{i}\otimes K^i_b\rho_{ab}K^{i\dagger}_b\otimes\ket{i}_{b_1}\bra{i} \right),
\end{align*}
where registers $M_a$ and $M_b$ store classical information in party $a$ and $b$.

In the end, similar to the first step, the local operation that Alice performed based on the classical information received from Bob will also not increase $\mathcal{E}_D$. In fact, her local operation can be realized by adding ancillary subsystem $a_1$ in the initial pure state $\ket{0}\bra{0}$, followed by a unitary operation on $aa_1$ and subsystem $a_1$ is traced out at last.

In conclusion, one has
\begin{align*}
&\mathcal{E}_D(\rho_{ab})\ge \mathcal{E}_D \left(\sum_iK^i_b\rho_{ab}K^{i\dagger}_b\otimes\ket{i}_{b_1}\bra{i} \right)\\
\ge& \mathcal{E}_D \left(\sum_i\ket{i}_{a_1}\bra{i}\otimes K^i_b\rho_{ab}K^{i\dagger}_b\otimes\ket{i}_{b_1}\bra{i} \right)\\
=&\mathcal{E}_D [U_{aa_1a_2}(\ket{0}_{a_2}\bra{0}\otimes\\
&\sum_i\ket{i}_{a_1}\bra{i}\otimes K^i_b\rho_{ab}K^{i\dagger}_b\otimes\ket{i}_{b_1}\bra{i})U^{\dagger}_{aa_1a_2}]\\
\ge&\mathcal{E}_D \left(\sum_{i,j}\ket{i}_{a_1}\bra{i}\otimes K^{i,j}_aK^i_b\rho_{ab}K^{i\dagger}_bK^{i,j\dagger}_a\otimes\ket{i}_{b_1}\bra{i} \right).
\end{align*}
The last line implies that when Alice performs an operation conditioned on the
classical information received from Bob, the measure does not increase. This completes the proof.

\end{proof}

\section{Proof of Lemma \ref{convexity-induced-Bures-entanglement}}\label{convex-of-CSE-Bures-discord}

We have to prove the following lemma.

{\em Lemma}: $\mathcal{E}_{D^G_B}$ is convex in the sense that
\begin{align*}
\mathcal{E}_B \left(\sum_ip_i\rho^i_{ab} \right) \le\sum_ip_i\mathcal{E}_B(\rho^i_{ab}),
\end{align*}
where $p_i$ are probabilities and $\rho^i_{ab}$ are bipartite states shared
by parties $a$ and $b$.

\begin{proof}
Note that
\begin{align*}
\rho_{aa^{\prime}a^{\prime\prime}bb^{\prime}b^{\prime\prime}}:=\sum_ip_i\rho^i_{aa^{\prime}bb^{\prime}}\otimes\ket{i,i}_{a^{\prime\prime}b^{\prime\prime}}\bra{i,i}
\end{align*}
is a CSE of $\rho_{ab}=\sum_ip_i\rho^i_{ab}$ whenever $\rho^i_{aa^{\prime}:bb^{\prime}}$ is a CSE of $\rho^i_{ab}$ for all $i$. Suppose there exists a unitary $U^i_{aa^{\prime}}$ such that $U^i_{aa^{\prime}}\rho^i_{ab}\rho^{i\dagger}_{ab}$ is invariant under the SWAP operation between $a(b)$ and $a^{\prime}(b^{\prime})$. Then it is easy to check that the unitary $U_{aa^{\prime}a^{\prime\prime}}:=\sum_iU^i_{aa^{\prime}}\otimes\ket{i}_{a^{\prime\prime}}\bra{i}$ leaves $U_{aa^{\prime}a^{\prime\prime}}\rho_{aa^{\prime}a^{\prime\prime}bb^{\prime}b^{\prime\prime}}U^{\dagger}_{aa^{\prime}a^{\prime\prime}}$ invariant under the SWAP operation between $a(b)$ and $a^{\prime}(b^{\prime})$.

Without loss of generality, suppose $\rho^i_{aa^{\prime}:bb^{\prime}}$ is the optimal state extension of each $\rho^i_{ab}$ and $\sigma^{\star}_i$ is the corresponding closest classical correlated state. Then
\begin{align*}
&\sum_ip_i\mathcal{E}_{D^G_B}(\rho^i_{ab})=\sum_ip_id^2_B \left(\rho^i_{aa^{\prime}bb^{\prime}},\sigma^{\star}_i \right)\nonumber\\
=&\sum_ip_id^2_B \left(\rho^i_{aa^{\prime}bb^{\prime}}\otimes\ket{i,i}_{a^{\prime\prime}b^{\prime\prime}}\bra{i,i},\sigma^{\star}_i\otimes\ket{i,i}_{a^{\prime\prime}b^{\prime\prime}}\bra{i,i} \right)\nonumber\\
\ge & d^2_B \left(\sum_ip_i\rho^i_{aa^{\prime}bb^{\prime}}\otimes\ket{i,i}_{a^{\prime\prime}b^{\prime\prime}}\bra{i,i},\sum_ip_i\sigma^{\star}_i\otimes\ket{i,i}_{a^{\prime\prime}b^{\prime\prime}}\bra{i,i} \right)\nonumber\\
\ge&D^G_B \left(\sum_ip_i\rho^i_{aa^{\prime}bb^{\prime}}\otimes\ket{i,i}_{a^{\prime\prime}b^{\prime\prime}}\bra{i,i} \right)\nonumber\\
\ge&\mathcal{E}_{D^G_B} \left(\sum_ip_i\rho^i_{ab} \right),
\end{align*}
where the first inequality follows from the joint convexity of $d^2_B$ and the second inequality is based on the definition of $\mathcal{E}_B$. The last inequality follows from the fact that $\sum_ip_i\rho^i_{aa^{\prime}bb^{\prime}}\otimes\ket{i,i}_{a^{\prime\prime}b^{\prime\prime}}\bra{i,i}$ is a CSE of $\sum_ip_i\rho^i_{ab}$.
\end{proof}


\begin{thebibliography}{99}
\bibitem{note2}If $d$ is such that $d(\rho,\sigma)=d(\rho\otimes\ket{0}\bra{0},\sigma\otimes\ket{0}\bra{0})$, then the corresponding geometric discord $D^G_d$ and MID $D^M_d$ are both non-increasing when adding a pure state ancilla.
Firstly, assume $\sigma^{\star}_{ab}$ is the closest classical correlated state, then
\begin{align*}
&D^G_d(\rho_{ab})=d(\rho_{ab},\sigma^{\star}_{ab})\\
=&d(\rho_{ab}\otimes\ket{00}_{a_1b_1}\bra{00},\sigma^{\star}_{ab}\otimes\ket{00}_{a_1b_1}\bra{00})\\
\ge&\min_{\sigma_{ab}\in\mathcal{C}\mathcal{C}}d(\rho_{ab}\otimes\ket{00}_{a_1b_1}\bra{00},\sigma_{ab})=D^G_d(\rho_{ab}\otimes\ket{00}_{a_1b_1}\bra{00}).
\end{align*}
Secondly, assume $\Pi^{\star}$ is the optimal measurement, then
\begin{align*}
&D^M_d(\rho_{ab})=d(\rho_{ab},\Pi^{\star}(\rho_{ab}))\\
=&d(\rho_{ab}\otimes\ket{00}_{a_1b_1}\bra{00},\Pi^{\star}(\rho_{ab})\otimes\ket{00}_{a_1b_1}\bra{00})\\
\ge&\min_{\Pi}d(\rho_{ab}\otimes\ket{00}_{a_1b_1}\bra{00},\Pi(\rho_{ab}\otimes\ket{00}_{a_1b_1}\bra{00}))\\
=&D^M_d(\rho_{ab}\otimes\ket{00}_{a_1b_1}\bra{00}).
\end{align*}
\end{thebibliography}
\end{document}